\documentclass{article}

\begin{document}

\title{Random Walks and Effective Resistances on Toroidal and Cylindrical Grids}
\author{Monwhea  Jeng \\
momo@physics.ucsb.edu \\
Physics Department \\
University of California, Santa Barbara, CA 93106-4030}
\maketitle

\begin{abstract}
\noindent A mapping between random walk problems and resistor 
network problems is described and used to calculate the effective 
resistance between any two nodes on an infinite two-dimensional
square lattice of unit resistors. The superposition principle is 
then used to find effective resistances on toroidal
and cylindrical square lattices.
\end{abstract}


\section{Introduction}

There is an interesting but little-known correspondence between 
properties in random walk problems and properties in electric
network problems~\cite{Doyle}. In this paper we describe this
correspondence and show how it can be used to calculate
resistances between arbitrary nodes on an infinite two-dimensional 
square lattice of unit resistors. While this problem has been solved
elsewhere~\cite{Venezian}-~\cite{Atkinson}, the treatment here both shows
the value of mapping electric network problems to random 
walk problems, and puts the answer in a form that can, by use of the
superposition principle, be used to calculate resistances on toroidal 
and cylindrical grids.


\section{Random Walks and Effective Resistances}
\label{sec-map}

In this section we demonstrate a number of surprising
relationships between resistor networks
and certain random walk problems. A very lucid explanation of the results
covered here, as well as other aspects of this mapping, can be
found in~\cite{Doyle}.

We first consider a general finite connected resistor network (Fig. 1).
If $x$ and $y$ are connected nodes, let the resistor connecting them have
resistance $r_{xy}$. We now consider a random walker who goes from
site to site, weighing each possible step by its inverse resistance.
To be specific, if $N(x)$ is the set of all nodes connected to $x$
by a single resistor, then the probability that a random walker
at $x$ will next move to the node $y\in N(x)$ is

\begin{equation}
p_{x\rightarrow y} = \frac{1}{c_x r_{xy}}\ ,\ \ \ \mathrm{where}\ 
c_x\equiv\sum_{y\in N(x)} \frac{1}{r_{xy}}
\end{equation}

Now put nodes $A$ and $B$ at voltages 1 and 0, and let current flow
through the network, with no sources at nodes besides $A$ and $B$. 
Then $V_x$, the voltage at an arbitrary point $x$, can be
interpreted as the probability that the above random walker,
starting at $x$, will get to $A$ before $B$.
To see this, we first note that this probability interpretation 
clearly works at the boundary conditions
$V_A=1$ and $V_B=0$. At other points, $y\neq A$ or $B$, there
is no current source, so from Kirchoff's laws,

\begin{eqnarray}
\nonumber
0 & = & \sum_{y\in N(x)} I_{x\rightarrow y} = 
\sum_{y\in N(x)} \frac{V_x-V_y}{r_{xy}} =
V_x\sum_{y\in N(x)}\frac{1}{r_{xy}} - \sum_{y\in N(x)}\frac{V_y}{r_{xy}} =\\
 & = & c_x (V_x-\sum_{y\in N(x)}p_{x\rightarrow y}V_y)
\end{eqnarray}

\noindent And $V_x=\sum_{y\in N(x)}p_{x\rightarrow y}V_y$ is exactly the 
relationship that we would write down for the
probability $V_x$ that a random walker starting at $x$ 
would reach $A$ before $B$. Since both the resistor and random walk
problems have the same boundary conditions and solve the same linear
equations, they have the same unique solution (although, technically, 
for an infinite lattice the solution is not unique -- see 
section~\ref{sec-inf} for more details). 

We now calculate the current from $A$ to $B$:

\begin{eqnarray}
\nonumber
I & = & \sum_{y\in N(A)} I_{A\rightarrow y} =
\sum_{y\in N(A)} \frac{V_{Ay}}{r_{Ay}} =
c_A \sum_{y\in N(A)} p_{A\rightarrow y} (1-V_y) = \\
\nonumber
 & = & c_A \sum_{y\in N(A)} p_{A\rightarrow y} \times
(\mathrm{probability}\ \mathrm{that}\ \mathrm{a}\ \mathrm{random}\ 
\mathrm{walker}\ \mathrm{at}\ y\ \mathrm{gets}\ \mathrm{to}\ B \ 
\mathrm{before}\ A) \\
\label{eq:current}
 & = & c_A p_{AB}
\end{eqnarray}

\noindent where we have used the random walk mapping,
and defined $p_{AB}$ as the probability that a random walker,
starting at $A$, gets to $B$ before returning to $A$.



The voltage between $A$ and $B$ is 1, and the current is given by
equation~\ref{eq:current}, so from Ohm's law, the effective resistance
between $A$ and $B$ is

\begin{equation}
\label{eq:randomwalk}
R_{AB}=\frac{1}{c_A p_{AB}}
\end{equation}

It will be useful to write this result in a different form.
For a random walker starting at A, let $\Delta_{AB}$ be the
expectation value of the number of vists to A minus the number of
vists to B, after infinitely many steps. If $P_n(x)$ is the probability
that after n steps the walker will be at node x, then 

\begin{equation}
\Delta_{AB} = \sum_{n=0}^{\infty} (P_n (A)-P_n (B))
\end{equation}

\noindent It is not hard to show from the definition of $\Delta_{AB}$,
that $\Delta_{AB}=1/2p_{AB}$, and thus

\begin{equation}
R_{AB}=\frac{2}{c_A}\Delta_{AB}=
\frac{2}{c_A}\sum_{n=0}^{\infty} (P_n (A)-P_n (B))
\end{equation}




\section{$R_{eff}$ on an Infinite Grid}
\label{sec-inf}

In this section we show how the random walk mapping can be used to
find effective resistances on an infinite two-dimensional grid of
unit resistors. This problem has been solved 
elsewhere~\cite{Venezian}--~\cite{Atkinson}, but we rederive the result 
here to demonstrate the power of the mapping described above.

We can solve the corresponding random walk problem with a
generating function~\cite{Spitzer}. Let the random walker start
at position $(0,0)$. After $N$ timesteps she is at position
$\vec{x}_{N} = \sum_{i=1}^{N} \hat{e}_i$, where $\hat{e}_{i}$ is the
step at timestep $i$, and each $\hat{e}_{i}$ is chosen with 
equal probability from $(0,1)$, $(0,-1)$, $(1,0)$, and $(-1,0)$. 
Then the expectation value of $e^{i \hat{e} \cdot \vec{\theta} }$, where
$\hat{e}$ is any step, and $\vec{\theta}$ is a 2-vector, is

\begin{eqnarray}
\lefteqn{ \hspace{-0.52in}
\phi(\vec{\theta}) \equiv E(e^{i\hat{e}\cdot\vec{\theta}})
	=\frac{1}{2}(\cos\theta_x+\cos\theta_y) \;\;\rm{,while}   }\\
E(e^{i \vec{x}_{N} \cdot \vec{\theta} }) & = &
	E( \prod_{i=1}^{N} e^{i\hat{e}_i\cdot\vec{\theta} }) =
	\left[ \phi(\vec{\theta}) \right]^N =
	\phi^N (\theta)
\end{eqnarray}

\noindent Fourier tranforming, the probability of being at
$\vec{x}$ at timestep $N$ is

\begin{equation}
P_N (\vec{x}) = E(\delta_{\vec{x},\vec{x}_N}) =
	\frac{1}{(2\pi)^2} 
	\int_{-\pi}^{\pi} d\theta_x \int_{-\pi}^{\pi} d\theta_y
	e^{-i \vec{x} \cdot \vec{\theta} }
	\phi^N(\vec{\theta})
\end{equation}

\noindent Let $\Delta_{mn}^{\infty\infty} \equiv 
\Delta_{(0,0),(m,n)}$. The ``$\infty\infty$'' superscript indicates
that the grid is infinite in both length and width, and 
$\Delta_{(0,0),(m,n)}$ was defined in the last section.

\begin{eqnarray}
\nonumber
\Delta_{mn}^{\infty\infty} & = &
\sum_{N=0}^{\infty} (P_N (0,0)-P_N (m,n)) \\
\nonumber
 & = &  \frac{1}{(2\pi)^2} 
	\int_{-\pi}^{\pi} d\theta_x \int_{-\pi}^{\pi} d\theta_y
	(1-e^{-i (m,n) \cdot \vec{\theta} } )
	\sum_{N=0}^{\infty} \phi^{N}(\vec{\theta}) \\
\label{eq:Delmninf}
 & = & \frac{1}{(2\pi)^2}
	\int_{0}^{2\pi} d\theta_x \int_{0}^{2\pi} d\theta_y
	\frac{ 1-e^{-i (m,n) \cdot \vec{\theta} } }
	     {1-\phi\vec{(\theta})} \\
\label{eq:Rmninf}
R_{mn}^{\infty\infty} & = & \frac{1}{8\pi^2}
	\int_{0}^{2\pi} dx \int_{0}^{2\pi} dy
	\frac{1-\cos (mx+ny)}{1-\frac{1}{2} (\cos x + \cos y)}
\end{eqnarray}

In the last line we have used the mapping
in section~\ref{sec-map} to turn the random walk
quantity $\Delta_{mn}^{\infty\infty}$ into
$R_{mn}^{\infty\infty}$, the effective resistance between
$(0,0)$ and $(m,n)$. We can get $R_{mn}^{\infty\infty}$
in closed form for any $(m,n)$. We find
$R_{01}^{\infty\infty}=\frac{1}{2}$,
either by evaluating the integral above, or more simply, by
exploiting the symmetry of the original 
problem~\cite{Bartis,Aitchison}. For $m=n$ we can evaluate the integral
exactly~\cite{Spitzer,FAQ,Cameron}, getting 
$R_{mm}^{\infty\infty}=\frac{2}{\pi}\sum_{i=1}^{m}\frac{1}{2i-1}$.
From these values of $R_{mn}^{\infty\infty}$, we can use 
the recursion relation
$R_{m,n+1}^{\infty\infty}+R_{m,n-1}^{\infty\infty}+
R_{m+1,n}^{\infty\infty}+R_{m-1,n}^{\infty\infty}=
4R_{m,n}^{\infty\infty}$ for $(m,n)\neq (0,0)$ (easily
derivable from equation~\ref{eq:Rmninf}), to get an exact
expression for any $R_{mn}^{\infty\infty}$.
As we will see in the next section, the above integral form of 
$R_{mn}^{\infty\infty}$ is useful for calculating
effective resistances on toroidal grids.


If we wish to be rigorous, we should note that for an infinite resistor 
network, Kirchoff's laws do not have a
unique solution. They do however have a unique
physical solution, obtainable by requiring that the total power
dissipitated be finite and that the current flow be the limit of a 
sequence of flows contained in finite subnetworks. A rigorous
theory of flows in general infinite networks can be found 
in~\cite{FlandersI,ZemanianInfinite}, while analyses specific for
the infinite square lattice can be found in~\cite{FlandersII,Cameron}.


\section{$R_{eff}$ for a Toroidal Grid}
\label{sec-tor}

With the solution to the infinite grid, we now turn out attention
to the new problem of a toroidal grid of unit resistors. 
We let the toroidal grid
be $M$ by $N$, and want to find $R_{mn}^{MN}$, the 
effective resistance between nodes $(0,0)$ and $(m,n)$.

First imagine inserting $(1-\frac{1}{MN})$ amps at $(0,0)$, and drawing out 
$\frac{1}{MN}$ amps at every other node (Fig. 2). Let $S_{mn}^{MN}$
be the voltage between $(0,0)$ and $(m,n)$ in this
set-up. The set-up in which $(1-\frac{1}{MN})$ amps 
is drawn out at $(m,n)$,
and $\frac{1}{MN}$  amps are inserted at all other nodes 
will, by symmetry, also have voltage $S_{mn}^{MN}$ 
between $(0,0)$ and $(m,n)$. Superimposing these two
solutions, we find that if we insert $1$ amp at $(0,0)$, and take
out $1$ amp at $(m,n)$, we will have voltage $2 S_{mn}^{MN}$
between $(0,0)$ and $(m,n)$, and thus $R_{mn}^{MN}=2 S_{mn}^{MN}$.

Similar reasoning tells us that for an
infinite grid, if we insert $1$ amp at $(0,0)$ and let
it escape to infinity, then the voltage difference between
$(0,0)$ and $(m,n)$ will be $R_{mn}^{\infty\infty}/2$~\cite{Venezian}.

We can now calculate $S_{mn}^{MN}$.
Because of the periodicity of the $M\times N$ toroidal grid, the
voltage drops on the torus when $(1-\frac{1}{MN})$ amps are
inserted at $(0,0)$
and $\frac{1}{MN}$ amps are drawn out at all other nodes, are the same as the
voltage drops on the infinite grid when $(1-\frac{1}{MN})$ amps are inserted at 
$(aM,bN)$ for all integers $a$ and $b$, and $\frac{1}{MN}$  amps are drawn out
at all other nodes. So instead of having the left and right
ends (and the top and bottom) wrap around in Fig. 2, we have them
repeat. We thus define

\begin{equation}
\label{eq:Iab}
I_{ab} \equiv \left\{
\begin{array}{cc} 1-\frac{1}{MN} & 
	\mathrm{if }\;\frac{a}{M}\,\;\mathrm{and }\,\;\frac{b}{N}\,\;
		\mathrm{are}\ \mathrm{both}\ \mathrm{integers} \\
	-\frac{1}{MN} &
	\mathrm{otherwise}
\end{array} \right.
\end{equation}

\noindent as the current into site $(a,b)$. Each $I_{ab}$ induces a
voltage $I_{ab}(R_{a-m,b-n}^{\infty\infty}/2)$ at site $(m,n)$, and
a voltage $I_{ab}(R_{ab}^{\infty\infty}/2)$ at site
$(0,0)$. Superimposing these solutions, we get

\begin{equation}
\label{eq:superposition}
R_{mn}^{MN} = 2 S_{mn}^{MN} =
	\sum_{a=-\infty}^{\infty} \sum_{b=-\infty}^{\infty} 
	I_{ab} (R_{(a,b),(m,n)}^{\infty\infty} -
		R_{(a,b),(0,0)}^{\infty\infty} )
\end{equation}

\noindent Equations~\ref{eq:Iab} and~\ref{eq:superposition} contain all
the physics. The rest is just mathematical manipulation. 

\begin{eqnarray}
\nonumber
R_{mn}^{MN} & = & \frac{1}{8\pi^2} 
	\sum_{a=-\infty}^{\infty} \sum_{b=-\infty}^{\infty}
	I_{ab} \int_{0}^{2\pi} dx \int_{0}^{2\pi} dy \\
\nonumber
\lefteqn{  \hspace{1.0in}
	\frac{\cos (ax+by) - \cos ((a-m)x+(b-n)y)}
	{1-\frac{1}{2}(\cos x + \cos y)}  } \\
\nonumber
 & = & \frac{1}{8\pi^2} \int_{0}^{2\pi} dx \int_{0}^{2\pi} dy
	\frac{1-\cos (mx+ny)}{1-\frac{1}{2}(\cos x + \cos y)} \\
\lefteqn{  \hspace{1.0in}
	\left[ \sum_{a=-\infty}^{\infty} \sum_{b=-\infty}^{\infty}
	I_{ab} \cos (ax) \cos (by) \right] }
\label{eq:RmnMNa}
\end{eqnarray}

\noindent We can do the sums over $a$ and $b$ exactly, using the 
following identity :

\begin{eqnarray}
\nonumber
\sum_{a=-\infty}^{\infty} \cos (aKx) & = &
   \lim_{p\rightarrow\infty} \sum_{a=-p}^{p} 
   \left( e^{iKx} \right)^a =
   \lim_{p\rightarrow\infty} 
   \frac{\sin ((p+\frac{1}{2})Kx)}{\sin (\frac{1}{2}Kx)} \\
\lefteqn{\hspace{-0.5in}  = 
   \frac{2\pi}{K} \sum_{u=-\infty}^{\infty} 
   \delta (x-\frac{2\pi}{K}u) }
\end{eqnarray}

\noindent Here we first did the geometric sum exactly, and then used the
representation of the Dirac delta function,
$\lim_{p\rightarrow\infty} \frac{\sin (pz)}{z} = \pi\delta (z)$.
Using this result, we get

\begin{eqnarray}
\nonumber
\lefteqn{
 \hspace{-2.2in} 
 \sum_{a=-\infty}^{\infty} \sum_{b=-\infty}^{\infty} 
     I_{ab} \cos (ax) \cos (by) } \\
\lefteqn{
 \hspace{-1.9in}
 = \frac{4\pi^2}{MN} 
	\sum_{u=-\infty}^{\infty} \sum_{v=-\infty}^{\infty}
	\left[ \delta (x-\frac{2\pi}{M}u) \delta (y-\frac{2\pi}{N}v) -
		\delta (x-2\pi u) \delta (y-2\pi v) \right]
}
\end{eqnarray}

\noindent Inserting this back into equation~\ref{eq:RmnMNa}, we can 
immediately do the integrals over $x$ and $y$, getting

\begin{equation}
\label{eq:RmnMNb}
R_{mn}^{MN} = \frac{1}{2MN} 
	\sum_{u=0}^{M-1} \sum_{v=0}^{N-1} {}^{\prime}
	\frac{1-\cos (2\pi (m\frac{u}{M}+n\frac{v}{N}) )}
          {1-\frac{1}{2} (\cos (2\pi\frac{u}{M}) + \cos (2\pi\frac{v}{N}))}
\end{equation}

\noindent where the prime on the sum indicates that we omit
the term $(u,v)=(0,0)$. We note that this formula
immediately implies that $R_{01}^{MN}+R_{10}^{MN}=1-\frac{1}{MN}$.


\section{$R_{eff}$ on a Cylindrical Grid}

We can find the results for an infinite cylindrical grid by
taking one the of the toroidal lengths to infinity. One of the sums
then becomes a Riemannian representation of an integral. For example,
if $M\rightarrow\infty$ we get

\begin{equation}
R_{mn}^{\infty N} = \frac{1}{4\pi N} \int_{0}^{2\pi} dx
	\sum_{v=0}^{N-1} 
	\frac{1-\cos (mx+2\pi n\frac{v}{N})}
	{1-\frac{1}{2}(\cos (x)+\cos (2\pi \frac{v}{N}))}\ \ ,
\end{equation}

\noindent which is ``halfway between''
equations~\ref{eq:Rmninf} and~\ref{eq:RmnMNb}. The
integral over $x$ can be done by contour
integration. For example, for $(m,n)=(0,1)$, we use 
$\int_{0}^{2\pi} \frac{dx}{k-\cos (x)} = \frac{2\pi}{\sqrt{k^2-1}}$
to get

\begin{equation}
R_{01}^{\infty N} = \frac{1}{N} \sum_{v=1}^{N-1}
	\sqrt{\frac{1-\cos (2\pi\frac{v}{N})}{3-\cos (2\pi\frac{v}{N})}}
\end{equation}

Calculating $R_{01}^{\infty 2}=\frac{1}{2\sqrt{2}}$ is a simple
freshman physics problem which can be solved by setting up the
right recursion relation. But it is more
difficult to show that $R_{01}^{\infty 3}=\frac{2}{\sqrt{21}}$.
Also, note that by the comment at the end of section~\ref{sec-tor},
$R_{10}^{\infty N} = 1 - R_{01}^{\infty N}$. 


\section{Conclusions}
\label{sec-conc}

This mapping may be used on any number of resistor problems or random
walk problems. Since a resistor problem and its equivalent random
walk problem are essentially the same Dirichlet problem, neither 
framework is inherently simpler. However, certain manipulations may be 
more intuitive and physically meaningful in one framework than another.
For example, the common freshman physics problem of calculating 
effective resistances on a cube of 1$\Omega$ resistors is best approached
by exploiting the symmetry of the cube to join points of equal voltage.
Effective resistances on other Platonic solids may be calculated
by the same method, or by cleverly superimposing two easily solvable
flows~\cite{Steenwijk}. While the same manipulations are possible in the 
equivalent random walk problem, they are not intuitive, and most
physicists would find it easiest to solve a random walk problem on an
icosohedron by first mapping it to the equivalent resistor problem.

On the other hand, for infinite lattices, the direct solution of the
resistor network by separation of variables has no obvious physical
meaning; but in the random walk framework the generating function is
both physically meaningful and natural. The various infinite
lattices considered in~\cite{Atkinson} can be solved by
changing the generating function (and some prefactors) in
equation~\ref{eq:Delmninf}. (We note that exact values for
effective resistances between any two points of triangular or 
honeycomb lattices can be obtained from recursion relations 
in~\cite{Horiguchi}.)

Perhaps the greatest advantage of mapping infinite resistor lattices 
to random walks is that many difficult random walk problems have already
been solved and their solutions are easily accessible.
Suppose we wish to
calculate $\lim_{l^2+m^2+n^2\rightarrow\infty}R_{lmn}$,
the resistance between the origin and infinity for
a three-dimensional cubic lattice. The resulting integrals 
are extraordinarily difficult to evaluate. However, after
using the random walk mapping we get

\begin{eqnarray}
\lim_{l^2+m^2+n^2\rightarrow\infty}R_{lmn} & = &
	\frac{1}{3}E(\# \ \mathrm{of}\ \mathrm{vists}\ \mathrm{to} \ \vec{0} \ 
	\mathrm{of}\ \mathrm{a}\ \mathrm{random}\ \mathrm{walk}\ 
	\mathrm{starting}\ \mathrm{at}\ \vec{0}) \\
 & = & \frac{1}{16\pi^3\sqrt{6}}\Gamma(\frac{1}{24})\Gamma(\frac{5}{24})
	\Gamma(\frac{7}{24})\Gamma(\frac{11}{24})
   = 0.50546. . .
\end{eqnarray}

\noindent simply by copying results from the random walk 
literature~\cite{Watson,Glasser}.

\bigskip
\bigskip
\bigskip

This work was supported by a UC Regents Fellowship. I would like
to thank the referees and editors for pointing out numerous missed
references, and Kerry Kuehn for helpful comments.


\pagebreak


\pagebreak

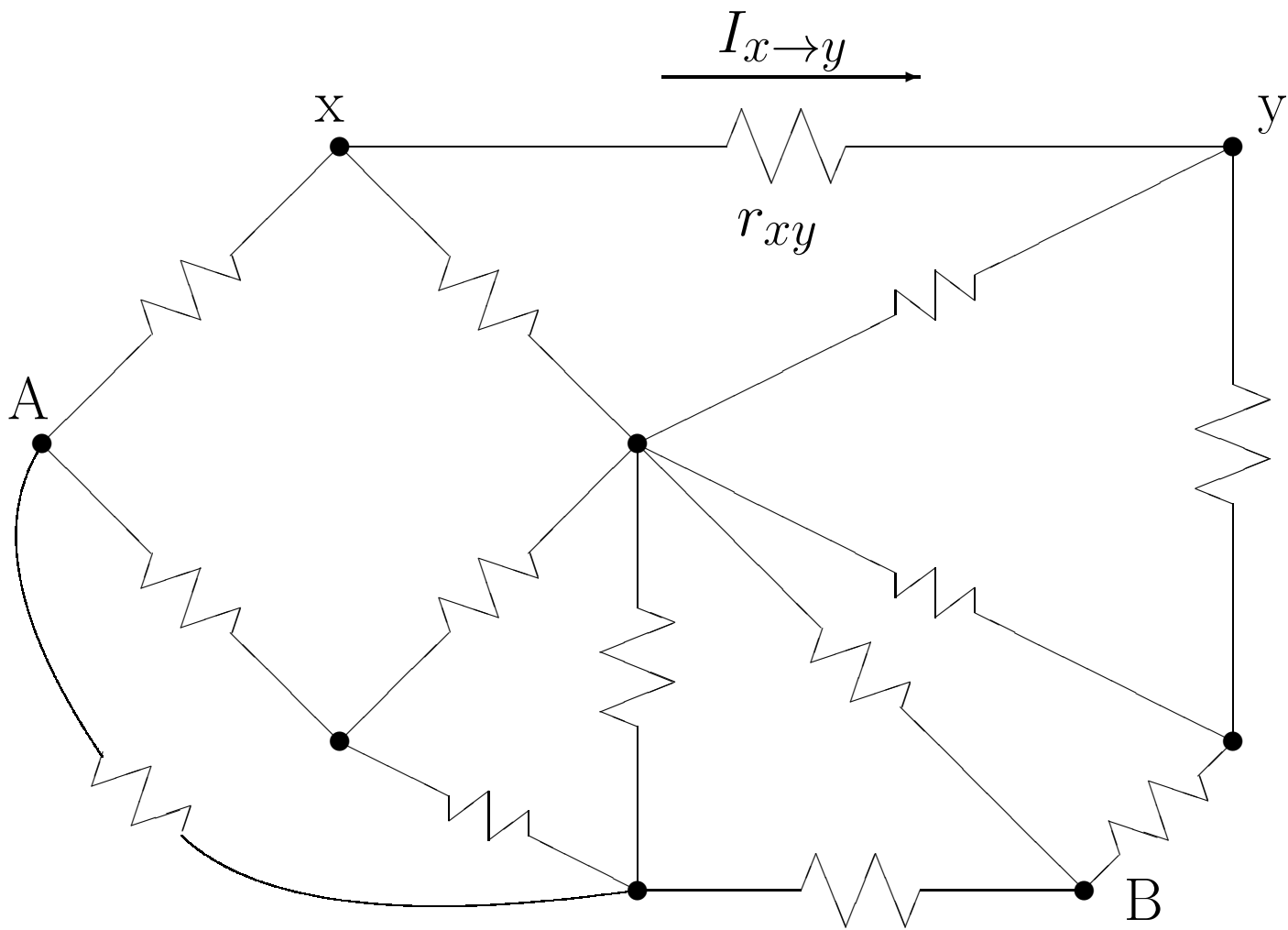
\noindent Figure 1 -- A generic resistor network

\bigbreak

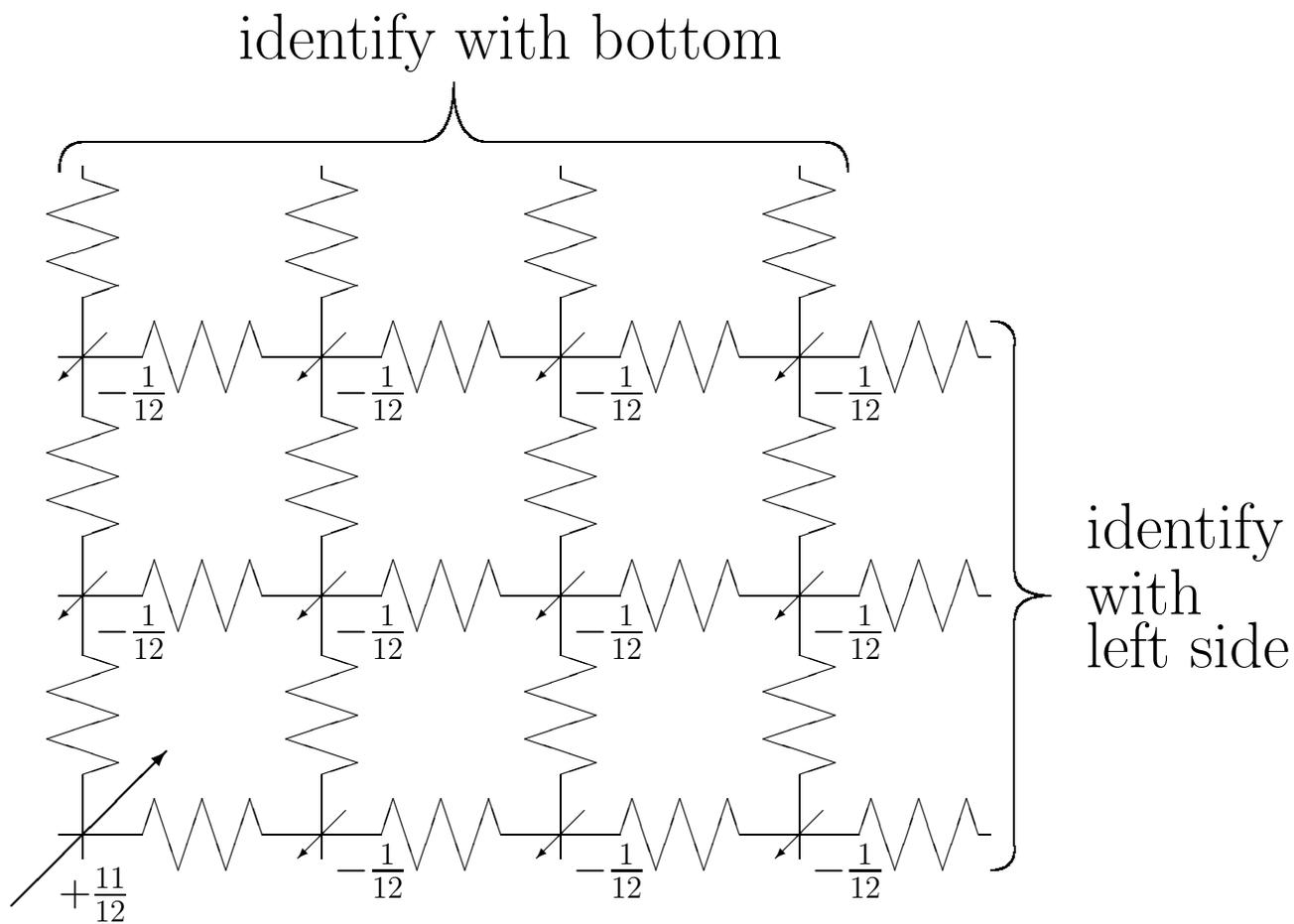
\noindent Figure 2 -- In and out currents on an M by N 
toroidal grid of unit resistors, for M=4, N=3


\pagebreak

\Huge


\setlength{\unitlength}{0.072cm}

\newsavebox{\hrescen}
\savebox{\hrescen}(0,0){
   \put(-12,0){\line(2,5){3}}
   \put(-9,7.5){\line(2,-5){6}}
   \put(-3,-7.5){\line(2,5){6}}
   \put(3,7.5){\line(2,-5){6}}
   \put(9,-7.5){\line(2,5){3}}
}

\newsavebox{\vrescen}
\savebox{\vrescen}(0,0){
   \put(0,12){\line(5,-2){7.5}}
   \put(-7.5,3){\line(5,2){15}}
   \put(-7.5,3){\line(5,-2){15}}
   \put(-7.5,-9){\line(5,2){15}}
   \put(-7.5,-9){\line(5,-2){7.5}}
}

\newsavebox{\upslpres}
\savebox{\upslpres}(0,0){
   \put(-10,-10){\line(1,1){2}}
   \put(-10,-2){\line(1,-3){2.3}}
   \put(-10,-2){\line(3,-1){12}}
   \put(-2,6){\line(1,-3){4}}
   \put(-2,6){\line(3,-1){12}}
   \put(8,8){\line(1,1){2}}
   \put(10,2){\line(-1,3){1.7}}
}

\newsavebox{\downslpres}
\savebox{\downslpres}(0,0){
   \put(-10,10){\line(1,-1){2}}
   \put(-10,2){\line(1,3){2.3}}
   \put(-10,2){\line(3,1){12}}
   \put(-2,-6){\line(1,3){4}}
   \put(-2,-6){\line(3,1){12}}
   \put(8,-8){\line(1,-1){2}}
   \put(10,-2){\line(-1,-3){1.7}}
}

\newsavebox{\tiltupres}
\savebox{\tiltupres}(0,0){
   \put(-10,-5){\line(2,1){2}}
   \put(-8,1){\line(0,-1){5}}
   \put(-8,1){\line(4,-3){8}}
   \put(0,5){\line(0,-1){10}}
   \put(0,5){\line(4,-3){8}}
   \put(8,4){\line(0,-1){5}}
   \put(8,4){\line(1,1){2}}
}

\newsavebox{\tiltdownres}
\savebox{\tiltdownres}(0,0){
   \put(-10,5){\line(2,-1){2}}
   \put(-8,-1){\line(0,1){5}}
   \put(-8,-1){\line(4,3){8}}
   \put(0,-5){\line(0,1){10}}
   \put(0,-5){\line(4,3){8}}
   \put(8,-4){\line(0,1){5}}
   \put(8,-4){\line(1,-1){2}}
}

\begin{picture}(300,200)(75,-50)

   \put(30,100){\circle*{4}}
   \put(90,160){\circle*{4}}
   \put(90,40){\circle*{4}}
   \put(150,10){\circle*{4}}
   \put(150,100){\circle*{4}}
   \put(240,10){\circle*{4}}
   \put(270,40){\circle*{4}}
   \put(270,160){\circle*{4}}

   \put(30,100){\line(1,1){22}}   \put(90,160){\line(-1,-1){22}}
   \put(30,100){\line(1,-1){22}}  \put(90,40){\line(-1,1){22}}
   \put(90,160){\line(1,-1){22}}  \put(150,100){\line(-1,1){22}}
   \put(90,40){\line(1,1){22}}    \put(150,100){\line(-1,-1){22}}
   \put(90,160){\line(1,0){78}}   \put(192,160){\line(1,0){78}}
   \put(150,100){\line(2,1){52}}  \put(270,160){\line(-2,-1){52}}
   \put(150,100){\line(2,-1){52}} \put(270,40){\line(-2,1){52}}
   \put(270,40){\line(0,1){48}}   \put(270,160){\line(0,-1){48}}
   \put(150,100){\line(1,-1){37}} \put(240,10){\line(-1,1){37}}
   \put(240,10){\line(1,1){7}}    \put(270,40){\line(-1,-1){7}}
   \put(150,10){\line(1,0){33}}   \put(240,10){\line(-1,0){33}}
   \put(150,10){\line(0,1){33}}   \put(150,100){\line(0,-1){33}}
   \put(90,40){\line(2,-1){22}}   \put(150,10){\line(-2,1){22}}

   \qbezier(30,100)(15,77)(42,37) \qbezier(58,21)(80,0)(150,10)

   \put(180,160){\usebox{\hrescen}}
   \put(195,10){\usebox{\hrescen}}
   \put(150,55){\usebox{\vrescen}}
   \put(270,100){\usebox{\vrescen}}
   \put(60,130){\usebox{\upslpres}}
   \put(120,70){\usebox{\upslpres}}
   \put(255,25){\usebox{\upslpres}}
   \put(60,70){\usebox{\downslpres}}
   \put(120,130){\usebox{\downslpres}}
   \put(50,30){\usebox{\downslpres}}
   \put(195,55){\usebox{\downslpres}}
   \put(210,130){\usebox{\tiltupres}}
   \put(210,70){\usebox{\tiltdownres}}
   \put(120,25){\usebox{\tiltdownres}}

   \put(23,105){A}
   \put(248,4){B}
   \put(275,165){y}
   \put(85,165){x}
   \put(170,142){$r_{xy}$}
   \put(166,179){$I_{x\rightarrow y}$}
   \thicklines
   \put(155,174){\vector(1,0){52}}
   \thinlines

   \normalsize
   \put(50,-100){\shortstack[l]{Figure 1\\Monwhea Jeng\\Random Walks and
Effective Resistances on Toroidal and Cylindrical Grids}}
\end{picture}


\setlength{\unitlength}{0.032cm}

\newsavebox{\hres}
\savebox{\hres}(100,50){
   \begin{picture}(100,50)
      \put(0,25){\line(1,0){25}}
      \put(25,25){\line(1,3){5}}
      \put(30,40){\line(1,-3){10}}
      \put(40,10){\line(1,3){10}}
      \put(50,40){\line(1,-3){10}}
      \put(60,10){\line(1,3){10}}
      \put(70,40){\line(1,-3){5}}
      \put(75,25){\line(1,0){25}}
   \end{picture} }

\newsavebox{\vres}
\savebox{\vres}(50,100){
   \begin{picture}(50,100)
      \put(25,0){\line(0,1){25}}
      \put(25,25){\line(3,1){15}}
      \put(40,30){\line(-3,1){30}}
      \put(10,40){\line(3,1){30}}
      \put(40,50){\line(-3,1){30}}
      \put(10,60){\line(3,1){30}}
      \put(40,70){\line(-3,1){15}}
      \put(25,75){\line(0,1){25}}
   \end{picture} }

\newsavebox{\hrestrunc}
\savebox{\hrestrunc}(100,50){
   \begin{picture}(100,50)
      \put(0,25){\line(1,0){25}}
      \put(25,25){\line(1,3){5}}
      \put(30,40){\line(1,-3){10}}
      \put(40,10){\line(1,3){10}}
      \put(50,40){\line(1,-3){10}}
      \put(60,10){\line(1,3){10}}
      \put(70,40){\line(1,-3){5}}
      \put(75,25){\line(1,0){5}}
   \end{picture} }

\newsavebox{\vrestrunc}
\savebox{\vrestrunc}(50,100){
   \begin{picture}(50,100)
      \put(25,0){\line(0,1){25}}
      \put(25,25){\line(3,1){15}}
      \put(40,30){\line(-3,1){30}}
      \put(10,40){\line(3,1){30}}
      \put(40,50){\line(-3,1){30}}
      \put(10,60){\line(3,1){30}}
      \put(40,70){\line(-3,1){15}}
      \put(25,75){\line(0,1){5}}
   \end{picture} }

\begin{picture}(500,400)(50,0)
   \multiput(25,0)(0,100){3}{\multiput(0,0)(100,0){3}{\usebox{\hres}}}
   \multiput(325,0)(0,100){3}{\usebox{\hrestrunc}}
   \multiput(15,25)(0,100){3}{\line(1,0){10}}
   \multiput(0,25)(100,0){4}{\multiput(0,0)(0,100){2}{\usebox{\vres}}}
   \multiput(0,225)(100,0){4}{\usebox{\vrestrunc}}
   \multiput(25,15)(100,0){4}{\line(0,1){10}}
   \put(90,350){identify with bottom}
   \put(445,95){\shortstack[l]{identify\\with\\left side}}
   \qbezier(15,303)(15,315)(25,315)
   \qbezier(165,315)(180,315)(180,340)
   \qbezier(180,340)(180,315)(195,315)
   \qbezier(335,315)(345,315)(345,303)
   \put(25,315){\line(1,0){140}}
   \put(195,315){\line(1,0){140}}
   \qbezier(405,10)(415,10)(415,20)
   \qbezier(415,115)(415,125)(430,125)
   \qbezier(415,135)(415,125)(430,125)
   \qbezier(405,240)(415,240)(415,230)
   \put(415,20){\line(0,1){95}}
   \put(415,135){\line(0,1){95}}

   \multiput(35,235)(100,0){4}{\vector(-1,-1){20}}
   \multiput(35,135)(100,0){4}{\vector(-1,-1){20}}
   \multiput(135,35)(100,0){3}{\vector(-1,-1){20}}
   \thicklines
   \put(-5,-5){\vector(1,1){65}}

   \LARGE
   \put(15,-5){$+\frac{11}{12}$}
   \multiput(30,205)(100,0){4}{$-\frac{1}{12}$}
   \multiput(30,105)(100,0){4}{$-\frac{1}{12}$}
   \multiput(130,5)(100,0){3}{$-\frac{1}{12}$}

   \normalsize
   \put(0,-150){\shortstack[l]{Figure 2\\Monwhea Jeng\\Random Walks and 
Effective Resistances on Toroidal and Cylindrical Grids}}
\end{picture}


\begin{thebibliography}{99}

\bibitem{Doyle} Peter G. Doyle and J. Laurie Snell, Random
	Walks and Electric Networks (Mathematics Association
	of America, 1984), chapter 3 
\bibitem{Venezian} Giuluo Venezian, ``On the resistance between
	two points of a grid,'' Am. J. Phys. 62(11), 1000-1004 (1994)
\bibitem{FlandersII} Harley Flanders, ``Infinite Networks: II--Resistance 
	in an Infinite Grid,'' J. Math. Anal. and Applications,
	40, 30-35 (1972)
\bibitem{FAQ} rec.puzzles archive at 
	http://einstein.et.tudelft.nl/\string~arlet/puzzles/sol.cgi/physics/resistors
\bibitem{Cameron} David Cameron, ``The Square Grid of Unit Resistors,''
	Math. Scientist, 11, 75-82 (1986)
\bibitem{Trier} P. E. Trier, ``An Electrical Resistance Network and
	its Mathematical Undercurrents,'' Inst. Math. and its Applications,
	21, 58-60, (Mar/Apr 1985); P. E. Trier, ``Correspondence,''
	Inst. Math. and its Applications, 22, 30-31 (Jan/Feb 1986)
\bibitem{ZemanianLattice} A. H. Zemanian, ``A Classical Puzzle: The
	Driving-Point Resistances of Infinite Grids,'' IEEE Circuits
	and Systems Magazine, 7-9 (Mar 1984)
\bibitem{Kac} B. van der Pol, ``The Finite-Difference Analogy of the
	Periodic Wave Equation and the Potential Equation,'' Appendix
	IV in Probability and Related Topics in Physical
	Sciences (Interscience Publishers, London, 1959) by M. Kac
\bibitem{Lavatelli} L. Lavatelli, ``The Resistive Net and 
	Finite-Difference Equations,'' Am. J. Phys. 40, 1246-1257 (1972)
\bibitem{Atkinson} D. Atkinson and F. J. van Steenwijk, ``Infinite
	Resistive Lattices,'' Am. J. Phys. 67, 486-492 (1999)
\bibitem{Bartis} F. J. Bartis, ``Let's Analyze the Resistance Lattice,''
	Am. J. Phys., 35, 354-355 (1967)
\bibitem{Aitchison} R. E. Aitchison, ``Resistance Between Adjacent
	Points of Liebman Mesh,'' Am. J. Phys., 32(7), 566 (1964)
\bibitem{Spitzer} Frank Spitzer, Principles of Random 
	Walk (Springer-Verlag, NY, 1976), 2nd ed. 
\bibitem{FlandersI} Harley Flanders, ``Infinite Networks: I -- Resistive
	Networks,'' IEEE Trans. Circ. Theory, CT-18 (3), 326-331 (1971)
\bibitem{ZemanianInfinite} Armen H. Zemanian, ``Infinite Electrical
	Networks,'' Proc. IEEE, 64(1), 6-17 (1974)
\bibitem{Steenwijk} F. J. van Steenwijk, ``Equivalent Resistors of
	Polyhedral Resistive Structures,'' Am. J. Phys. 66 (1), 
	90-91 (1998)
\bibitem{Horiguchi} T. Horiguchi, ``Lattice Green's Functions for
	the Triangular and Honeycomb Lattics,'' J. Math. Phys.,
	13(9) 1411-1419 (1972)
\bibitem{Watson} G. N. Watson, ``Three Triple Integrals,'' Quarterly
	J. Math., 10, 266-276 (1939)
\bibitem{Glasser} M. L. Glasser and I. J. Zucker, ``Extended Watson 
	Integrals for the Cubic Lattices,'' Proc. Natl. Acad. Sci. USA,
	74(5), 1800-1801 (1977); ``Lattice Sums'' in Theoretical
	Chemistry: Advances and Perspectives, Volume 5 (Academic
	Press, New York, 1980), p.67-139

\end{thebibliography}
\end{document}